# NSGAN: A Non-Dominant Sorting Optimisation-Based Generative Adversarial Design Framework for Alloy Discovery.

Z. Li[1], N. Birbilis[1,2] *

[1]College of Engineering, Computing and Cybernetics, The Australian National University, Acton, A.C.T., 2601, Australia.

[2]Faculty of Science, Engineering, and Built Environment, Deakin University, Waurn Ponds, VIC, 3216, Australia

nick.birbilis@deakin.edu.au, zhipeng.li@anu.edu.au

**Abstract**

The design and discovery of new materials is fundamental to advancing scientific and technological innovation. The recent emergence of the materials genome concept holds great promise in revolutionising materials science by enabling the systematic utilisation of data for efficient prediction and optimisation of 'superior' materials. However, the materials genome approach can be stymied by the vast complexity of design spaces, which often demand substantial computational resources and sophisticated data processing capabilities. To address these challenges, this work introduces a novel generative design framework called the non-dominant sorting optimisation-based generative adversarial networks (NSGAN). Capitalising on the synergies of genetic algorithms (GA) and generative adversarial networks (GANs), NSGAN provides a robust and efficient approach for tackling high-dimensional multi-objective optimisation design problems. To validate the efficacy of the proposed framework, we applied the model to a comprehensive dataset of aluminium alloys. Additionally, an online tool was created as a supplementary resource, offering a brief introduction to this innovative method for the wider scientific community. This study explores the potential of a predictive and data-driven approach in material design, indicating a promising pathway for widespread applications in the field of materials science.

1. **Introduction**

The design and discovery of new materials has always been integral in propelling technological and scientific advances. The emergence of the concept of materials genome, analogous to biological genomes, holds substantial potential to revolutionise the field of materials science [1]. The materials genome approach emphasises the systematic production and organisation of data, which is then used to supply modelling and data science tools – for the ultimate goal of predicting, screening, and optimising materials with superior properties and reduced development time [2]. Computational approaches, such as density functional theory (DFT) and high-throughput screening (HTS), have emerged as valuable tools, enabling researchers to simulate material behaviour and rapidly evaluate a wide range of compositions and structures [3-6]. These methodologies allow for concurrent calculation of properties across a vast variety of materials; accelerating the pace of new materials discovery [6]. However, the comprehensive exploration of vast design spaces presents formidable challenges - as such methodologies are usually computationally expensive and place high demands on data processing and analytical capabilities.

The application of genetic algorithms in materials science has seen significant growth over the past few decades [7-9]. A genetical algorithm (GA) operates by mimicking the process of natural evolution, where it applies principles of survival of the fittest, crossover (breeding), and mutation to find solutions to a problem [10, 11]. By utilising a population-based stochastic searching approach, the GA inherently has the flexibility and adaptivity to explore vast solution space effectively - making it well-suited for solving complex design problems. The application of GA in materials design can be traced back to over two decades ago. One of the earliest applications of GA in materials science is found in the work of Ikeda, who utilised a GA for an efficient global search to optimise the equilibrium composition of multi-component alloys [8]. Dudiy and Zunger incorporated a GA in conjunction with the 'inverse band structure' approach, effectively searching for optimal alloy configurations based on pre-determined physical properties [12]. With the advent of machine learning (ML), there has been a trend towards integrating a GA with ML models in the field of materials design. In 2006, Anijdan *et al.* built a theoretical model using an artificial neural network (ANN) and a GA to optimise porosity formation in Al-Si casting alloys [13]. In that work, the ANN was used to predict the porosity percentage in Al–Si casting alloys, while the GA was utilised to search for optimal alloy designs. Similarly, Shen et al. devised a material design strategy that incorporated physical metallurgy guided ML models with GA - in the engineering of advanced ultrahigh-strength stainless steels [14]. More recently, Lee et al. proposed an inverse design methodology, utilising ML and GA, for the exploration and discovery of novel aluminium alloys optimised for enhanced ultimate tensile strength [15].

In the aforementioned applications, the GA is typically utilised to explore and optimise the design parameters of the potential candidates, with ML models tasked with predicting their intended properties. Owing to the GA's inherent flexibility and adaptivity in solving complex problems, this approach can be highly effective dealing with problems that possess a moderate number of parameters. Nonetheless, when addressing high-dimensional problems that have extensive search spaces, there remain inherent limitations with this approach. Since GA utilizes a stochastic searching based on a trial-and-error principle, the computational cost grows exponentially as the dimension of the problem increases. This becomes particularly problematic when the data being searched is distributed across a high-dimensional space (the 'design space') as a sparse manifold. When GA is applied directly to do optimisation in the design space, given that the initial population is randomly generated, it's highly probable that the searched samples significantly deviating from the data distribution of training data. Since ML models are generally trained on the training data, in such scenarios, the ML models' predictions for the searched samples can be unreliable. Moreover, erroneous predictions from the ML models regarding these samples could potentially lead the data distribution generated by the GA to never align with the distribution of training data. This eventually results in a state where the ML model's predictions remain unreliable indefinitely. Therefore, when tackling high-dimensional design problem, rather than conducting stochastic searching in the vast design space, it's more intuitive to initiate exploration and optimisation of novel samples near the data distribution of the training set. Given that existing data samples usually distribute nonlinearly in high-dimensional spaces, it is nearly impossible to locate the distribution of training data by simply limiting the range of each parameter (e.g., confining each element's molar ratio within a specific range). To address this, we propose a method that employs machine learning models to map this high-dimensional data distribution onto a simple, symmetric distribution. This mapping can then be utilised for direct exploration and optimisation on the distribution of training data.

Generative adversarial networks (GANs), first proposed by Goodfellow in 2014, have emerged as a potent machine learning tool for generating novel data samples [16]. The classic GAN architecture involves two neural networks - a generator and a discriminator. The generator is trained to generate synthetic data, while the discriminator's task is to determine whether these samples are real or generated [17]. A variant of GANs, the Wasserstein GANs (WGANs), employs the Earth Mover's distance as a measure of the discrepancy between the generator's distribution and the training data distribution, which enables the generator to learn to generate data that better approximates the original data distribution [18]. Since their inception, GANs have found diverse applications across various fields. Recently, the utility of GANs has begun to be recognised in the field of materials science [19-24]. For instance, one study proposed a framework based on convolutional neural networks (CNN) and GANs to inverse design the phase separation structure of polymer alloys and predict their Young's modulus [21]. Similarly, Xu and Hu proposed an efficient sampling method for the inverse design of metallic glasses based on Wasserstein GAN (WGAN) [22]. Some recent studies integrated an optimisation process within the GAN framework [24, 25]. In 2021, Long et al. put forth a constrained crystals deep convolutional generative adversarial network for the inverse design of crystal structures, where the generative model was constrained to generate distinct stable crystal structures through the optimisation of formation energy [24]. Abbasi et al. employed a feedbackGAN based optimisation strategy to force the generator towards generating samples with desired properties [25]. In these studies, given that the optimisation mechanism is implemented by imposing constraints on the generative network, the trained GAN models tend to generate biased data distributions. And one of the primary limitations of existing GAN-based approaches is their reliance on random sampling in the latent space, resulting in a majority of the generated samples bearing a strong resemblance to the training samples. Additionally, due to the highly nonlinear correlation between the latent and design spaces, it remains difficult to manipulate the model to generate samples with specific properties.

In this study, we proposed a novel generative design framework, non-dominant sorting optimisation-based generative adversarial networks (NSGAN), that can be applied in the exploration and discovery of novel materials with desired properties. The proposed generative design framework integrates the strengths of both GA and GANs in the design and optimisation of novel materials, compensating for the respective weaknesses of each model. When dealing with high-dimensional complex design problems, our approach leverages GANs to map the sparse high-dimensional data distribution onto a lower-dimensional, simple data distribution (in the latent space), addressing the so-called "curse of dimensionality" and significantly enhancing the search efficiency of the GA. By utilising the GA to do the searching and optimisation in the latent space, the GAN model is able to directionally generate novel samples with desired properties. Moreover, because the distribution of samples generated by the GAN model closely aligns with that of the original training data, when searching novel designs through the latent space, it substantially assures the predictive reliability of the property prediction ML models.

The efficiency and efficacy of NSGAN framework is demonstrated herein, utilising a dataset consisting of the composition, processing, and mechanical properties of 657 distinct aluminium (Al) alloys. Through the employment of a multi-objective optimisation evolutionary algorithm, our model aids in addressing typical conflicting objective optimisation problems in materials design, such as the strength-ductility trade-off in aluminium alloys. Additionally, we have developed an online tool that incorporates the proposed model, providing a platform for other researchers to leverage the design framework herein for their unique material design applications.

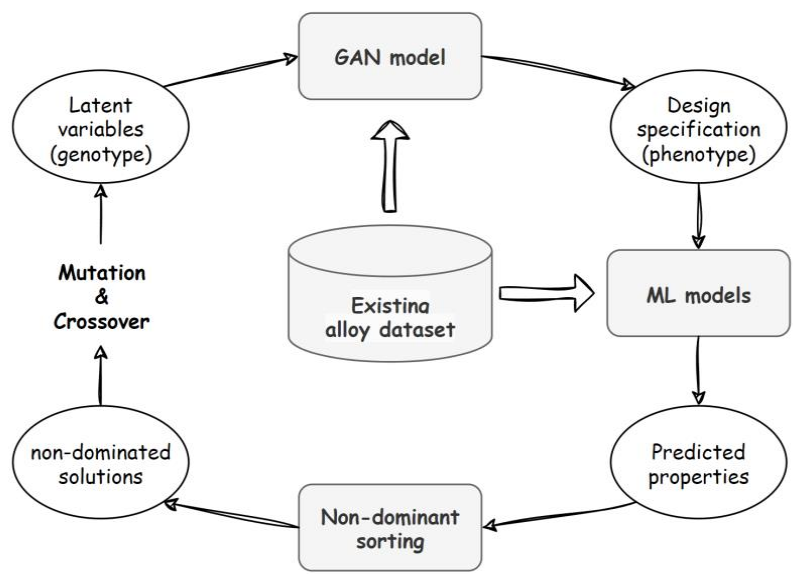

*Figure 1. Schematic representation of the proposed generative design framework.*

## 2. Alloy design strategy

The proposed non-dominant sorting optimisation-based generative adversarial design framework employs a method that integrates the capabilities of multi-objective optimisation algorithms with generative machine learning models, which primarily consists of three interwoven components: a GAN model, a series of ML models for property prediction, and a multi-objective optimisation evolutionary algorithm.

As depicted in Figure 1, the GAN model is trained to fit the distribution of the training dataset, including elemental composition, processing conditions, and (resultant) mechanical properties. In this procedure, the generator network of the GAN learns to map the distribution of the training data (the design space) onto a simple data distribution (the latent space, for which a multivariate Gaussian distribution is applied). In this framework, each data point in the latent space is treated as the genotype of a candidate alloy, and its corresponding output from the GAN model, representing the alloy's design specification, is considered its phenotype. It's worth noting that, in the context of evolutionary algorithms (EAs), the genotype serves as a concise (or encoded) representation of potential alloy designs, subjected to genetic operations such as crossover (recombination) and mutation. And the phenotype represents the observable characteristics of the actual solution that can be evaluated in the problem space. Effectively, the trained generator network of the GAN model functions as a transformer or a decoder for the candidate alloy, facilitating the mapping from genotype to phenotype.

The second component of our framework focuses on predicting the mechanical properties of the generated alloys using machine learning models. The dataset used for training encompasses various mechanical properties of aluminium alloys, namely yield strength, tensile strength, and elongation. To capture the nonlinear correlation between an alloy's design specification and its mechanical properties, we evaluated and compared the predictive performance of several widely applied ML models for regression, including random forests (RF) [26], gradient boosted tree (GBT) [27, 28], artificial neural networks (ANN) [29, 30], K-nearest Neighbours (KNN) [31], and support vector regression (SVR) [32, 33]. Among these, tree-based ensemble methods such as RF and GBT showcased the highest prediction accuracies and are therefore employed for the subsequent optimisation process.

In the optimisation procedure, our framework employs the non-dominated sorting genetic algorithm II (NSGA-II), which is a prominent evolutionary multi-objective optimisation algorithm [34]. The algorithm works by maintaining a population of potential solutions across generations. The genotype (the latent variables of the GAN) of the initial generation is derived from a multivariate Gaussian distribution, which results in the initial population as a random sampling of the distribution of training data. Utilising non-dominant sorting, a diverse set of non-dominated solutions is identified. And through processes of mutation and crossover, a new set of alloy candidates for the next generation is generated, which is subsequently transformed by the GAN model into the design specifications of candidate alloys. Leveraging the pre-trained ML model, the mechanical properties are then predicted and used as the optimisation objectives for the subsequent generation. The iterative optimisation process continues until a predefined stopping criterion is met, such as reaching a maximum number of generations or finding a solution that meets specific performance thresholds. Through this process, we aim to search for novel alloy candidates that potentially exhibit excellent mechanical properties, optimised to balance the strength-ductility trade-off.

### 3. Dataset

The Al alloy dataset employed in this study encompasses the characteristics of 657 distinct Al alloys, spanning a diverse compositional space across various series of aluminium alloys. The dataset includes alloys from all of the 1xxx to 8xxx series of Al alloys, encompassing a wide variety of alloy compositions. This is evident from the range, count, and average molar ratios of each alloying element present in the alloys, as detailed in the Table 1.

The element composition of the alloys is quantified by the molar ratios by the molar ratios of 25 elements present in the dataset, represented as $\mathbf{c} = [c_1, c_2, ..., c_{24}, c_{25}]^T$. This representation adheres to the constraint $\Sigma_{i=1}^{n} c_i = 1$, ensuring the sum of all molar ratios in any given alloy composition equals unity. Here, $c_i$ represents the molar ratio of the $i_{th}$ element, reflecting the proportional contribution of each element to the overall alloy composition.

*Table 1. Compositional range and molar ratios of Aluminium and top ten alloying elements in the dataset.*

| Element | Range | Count | Average Molar Ratio |
|---|---|---|---|
| Al | (0.75, 0.9999) | 657 | 0.941 |
| Mg | (0, 0.06) | 568 | 0.0164 |
| Cu | (0, 0.0685) | 478 | 0.0182 |
| Zn | (0, 0.12) | 357 | 0.0217 |
| Si | (0, 0.22) | 427 | 0.0178 |
| Mn | (0, 0.0131) | 432 | 0.00374 |
| Fe | (0, 0.012) | 399 | 0.00281 |
| Li | (0, 0.0382) | 45 | 0.0205 |
| Sn | (0, 0.2) | 4 | 0.131 |
| Cr | (0, 0.0411) | 300 | 0.00143 |
| Ti | (0, 0.0025) | 280 | 0.000793 |

In addition to elemental composition, the dataset also encompasses processing conditions and mechanical properties. As a crucial alloy descriptor that substantially impacts the final properties of alloys, based on the reported processing and heat treatment techniques, the processing conditions are incorporated as categorical features. Specifically, they are one-hot encoded into ten separate categories as follows: 1. Artificial aged, 2. Naturally aged, 3. No Processing, 4. Solutionised, 5. Solutionised + Artificially peak aged, 6. Solutionised + Artificially over aged, 7. Solutionised + Cold Worked + Naturally aged, 8. Solutionised + Naturally aged, 9. Strain hardened (Hard), 10. Strain hardened. The categories are meticulously defined in accordance with the standards set by the US Aluminium Association Alloy and Temper System [35].

The mechanical properties within the dataset, encompassing tensile strength, yield strength, and elongation, are comprehensively represented in the histograms shown in Figure 2. These histograms illustrate the frequency distributions for each mechanical property across all the alloys in the dataset. Each figure reflects a broad range of values, demonstrating the dataset's extensive variability and diversity.

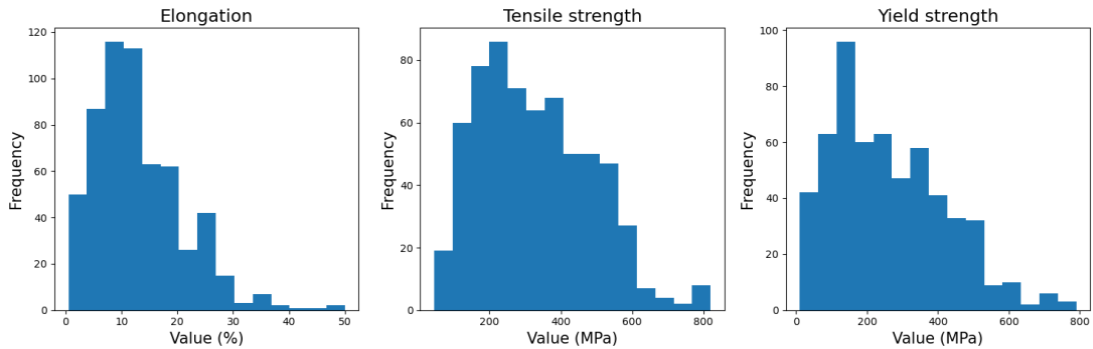

*Figure 2. Distribution histograms of mechanical properties in the alloy dataset.*

The dataset's comprehensive nature is demonstrated by the inclusion of a wide array of alloys, spanning multiple series, and diverse processing conditions, coupled with an extensive range of mechanical properties. This is also visually corroborated by the dimensionally reduced visualisation in Figure 3, where similar alloy samples are clustered together, indicating the diversity within the dataset. This diversity is critical for the robust training of machine learning models, enabling them to generalize effectively across diverse alloy compositions and mechanical characteristics.

To enable uniform treatment of these features and to facilitate an efficient training process, z-score normalization is employed. This procedure scales the features so that they possess a mean of 0 and a standard deviation of 1, thereby enhancing the model's overall performance and efficiency by ensuring scale sensitivity and accelerating convergence. The aluminium alloy dataset utilised in this research is open-access in the references [36, 37].

## 4. Property prediction models

Before delving into the generative and optimisation models, we will first introduce the ML models employed in property prediction. The application of ML models in predicting the mechanical properties of aluminium alloys has been investigated in several studies to date [15, 38-42]. Chaudry et al. employed various machine learning models to predict the hardness of Al-Cu-Mg-x alloys [39]. Their study demonstrated that the tree-based ensemble methods such as RF and GBT could achieve high performance in the prediction of the hardness of aluminium alloys. Li et al. demonstrated the feasibility of using ML to investigate the Al-Zn-Mg-Cu alloy system (7xxx series), highlighting its potential in expediting the discovery of high-performance alloys [43]. Similarly, Park et al. applied neural networks to predict the mechanical properties of 7xxx aluminium alloys. Assisted by an A.I.-based recommendation algorithm, new Al alloys were designed exhibiting a remarkable blend of strength and ductility [40].

In this study, we evaluated and compared the predictive performance of several widely applied nonlinear machine learning models for regression, including RF, GBT, ANN, KNN, and SVR. During the training process, we utilised 9-fold random subsampling cross-validation [44], and dividing the data into training, validation, and test sets at an 80:10:10 ratio. Hyperparameters tuning were performed using grid search for all the ML models, which is a method that systematically examines various combinations of hyperparameter tunes and performs cross-validation to determine the best model [45]. The prediction accuracies of the models were measured using the $R^2$ score, also known as the coefficient of determination, which provides a measure of how well the model replicate the variance of the observed values (true values) [46]. The formula for calculation is $R^2 = 1 - \frac{\Sigma(y_i - \hat{y}_i)^2}{\Sigma(y_i - \bar{y})^2}$, where $y_i$ is the observed value, $\hat{y}_i$ is the value predicted by the ML model, and $\bar{y}$ is the mean of the observed values in the dataset. From the formula it can be seen that the range of $R^2$ is $[-\infty, 1]$. An $R^2$ score of 1 is achievable only if every $y_i = \hat{y}_i$, which indicates that the ML model could predict the exact value for each test sample, thereby fully capturing the variance in the dependent variables. The predictive performance of various ML models for the mechanical property of the alloys in the database is shown in Table 2, which displays the models' average performance on the test set over 100 random splits of the datasets. This approach offers a robust representation of the models' generalization ability, effectively capturing their consistent performance across a diverse range of data subsets.

*Table 2.* The average and standard deviation of $R^2$ score, Mean Absolute Error (MAE), and Root Mean Squared Error (RMSE) for the prediction performance of mechanical properties using the machine learning models, where std denotes the standard deviation.

| Property | RF | | GBT | | ANN | | SVR | | KNN | |
|---|---|---|---|---|---|---|---|---|---|---|
| $R^2$ score | *mean* | *std* | *mean* | *std* | *mean* | *std* | *mean* | *std* | *mean* | *std* |
| Tensile strength | 0.917 | 0.025 | 0.922 | 0.023 | 0.887 | 0.045 | 0.899 | 0.034 | 0.889 | 0.037 |
| Yield strength | 0.878 | 0.043 | 0.889 | 0.040 | 0.839 | 0.055 | 0.863 | 0.047 | 0.862 | 0.045 |
| Elongation | 0.694 | 0.145 | 0.707 | 0.127 | 0.644 | 0.100 | 0.682 | 0.094 | 0.681 | 0.113 |
| **Error Metrics** | *MAE* | *RMSE* | *MAE* | *RMSE* | *MAE* | *RMSE* | *MAE* | *RMSE* | *MAE* | *RMSE* |
| Tensile strength (MPa) | 31.34 | 43.45 | 30.72 | 43.28 | 34.05 | 49.90 | 33.83 | 47.90 | 34.24 | 49.02 |
| Yield strength (MPa) | 36.71 | 53.19 | 36.64 | 52.72 | 40.42 | 58.96 | 38.30 | 56.56 | 37.80 | 56.94 |
| Elongation (%) | 2.92 | 4.24 | 2.87 | 4.06 | 3.46 | 4.60 | 3.22 | 4.40 | 3.26 | 4.45 |

It may therefore be observed (from Table 2), that both of the tree ensemble models, GBT and RF, outperform other methods in predicting the three properties of interest. Notably, GBT exhibits a marginal advantage over RF, attaining $R^2$ score accuracies of approximately 92.2% for tensile strength, 88.9% for yield strength, and 70.7% for elongation. In contrast, ANN although having comparable accuracy in predicting tensile strength, has noticeably lower accuracies in predicting yield strength and elongation compared to the other four methods. The possible underlying reason for this disparity in predictive performance among the models lies in their fundamental approach to modelling.

Unlike the other four methods listed in Table 2, the ANN is a parametric model that makes predictions by attempting to estimate the underlying functional form of the data. The output of ANN is the result of the input values processed through a nonlinear, complex mathematical formula. In contrast, non-parametric regression models, though varied in methodology, make fewer assumptions about the functional form, and rely primarily on the training data. For instance, KNN predicts based on the average target values of the k nearest neighbours in the feature space. Although this non-parametric nature makes them flexible and robust in handling complex relationships, they typically struggle to extrapolate outside the range of the training data. Parametric models, such as linear regression, polynomial regression, and ANN, could extrapolate values beyond the training data range more effectively when the chosen function aligns with the underlying data relationship. And this extrapolative ability becomes highly important in single-objective optimisation.

However, when the underlying relationship is complex, training a parametric model often requires abundant training data and meticulous consideration of the functional form. In our case, considering the high dimensionality and limited quantity of the given database, which complicates the training of a parametric model to fully capture the complex correlation and given that this study focuses on multi-objective optimisation, here we prioritize predictive performance over extrapolation ability. Therefore, GBT was selected the preferred model for property prediction in the remainder of this study.

## 5. Generative design models

The classic GAN typically consists of two subnetworks: a generative neural network G (the generator) and a discriminative neural network D (the discriminator). The input to the generator is a randomly distributed noise variable z, and its output is the synthetic (generated) data sample. The discriminator takes both real and synthetic samples as input, and outputs a value in the range [0, 1], activated by the sigmoid activation function. Here, 0 and 1 respectively signify the discriminator's judgement of the input data as fake or real.

Classic GAN models often suffer from issues like gradient vanishing and mode collapse (lack of diversity), leading to difficulty in training. Herein we employ the WGAN model which significantly mitigates these two issues. The WGAN model applies the discriminator as a critic to compute the Wasserstein distance (earth mover distance) between the generated and real data [18]. By employing the Wasserstein distance as the discriminative network's loss, the training of the generator is guided towards narrowing the overall discrepancy between the two datasets. Calculating this Wasserstein distance requires the discriminator network to fit a Lipschitz function, where the original WGAN applies weight clipping to restrict the neural network parameters within a range $[-c, c]$, where c is a positive small value, thus enforcing Lipschitz continuity. A further enhancement was made in the WGAN-GP (Gradient Penalty) paper, where weight clipping was abandoned in favour of adding a gradient penalty term to the discriminator network's loss [47]. The gradient penalty ensures that the gradient norms of the critic are close to 1, which aids in upholding the Lipschitz constraint necessary for the Wasserstein distance to work effectively. The authors demonstrated through comparisons across multiple databases that weight clipping could cause the WGAN to generate oversimplified data distributions, which is solve through using WGAN-GP.

To compare the performance of these two models in the generative design of aluminium alloys, both were used to generate 2000 samples, which were combined with the 657 training data points for t-SNE dimensional reduction and visualisation. As depicted in Figure 3, it may be observed that although both models can effectively approximate the distribution of the training data, the WGAN-GP-generated data show a closer resemblance to the actual training data distribution. In contrast, the WGAN model employing weight clipping generate a significant amount of data clustered around the training data or as interpolations of existing data. This comparison underscores the superiority of WGAN-GP in capturing the complex distribution of the dataset.

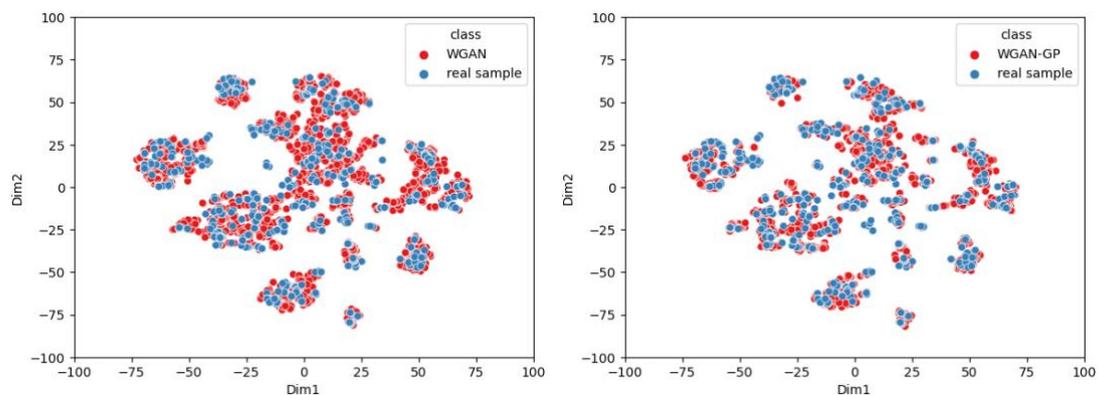

***Figure 3.*** *The t-SNE visualisation of the data distributions generated by the original WGAN model (left) and WGAN-GP model (right).*

Figure 4 depicts the architecture of the WGAN-GP model employed in this study, which consists of a generative neural network G (the generator) and a discriminative neural network D (the critic).

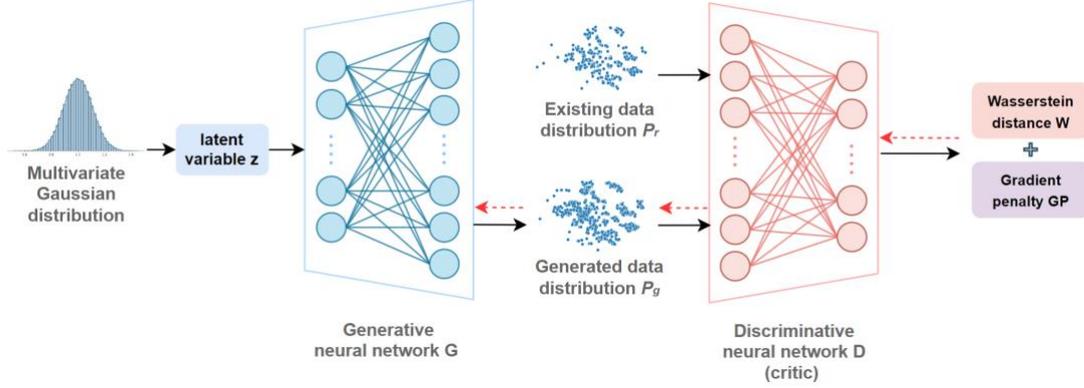

***Figure 4.*** *Schematic diagram of the WGAN-GP model employed herein. The generator takes a multivariate Gaussian distributed random noise z (latent variable) as input and is trained to learn the mapping to the data distribution of existing alloy samples. The path of error backpropagation is denoted by the red dashed arrows.*

Each of the subnetworks features two hidden layers, activated by LeakyReLU activation functions. A ReLU function serves as the output layer of the generator, ensuring that the generated data appear as sparse non-negative vectors. The input for the generator $\mathbf{z}$ is sampled from a 10-dimensional multivariate Gaussian distribution, $\mathbf{z} = [a_1, a_2, ..., a_9, a_{10}]^T$ where $a_i \sim N(0, 1)$. The loss function for the WGAN-GP model is shown as follows:

$$L = E_{\tilde{x} \sim P_g}[D(\tilde{x})] - E_{x \sim P_r}[D(x)] + \lambda E_{\hat{x} \sim P_{\hat{x}}}[(\|\nabla_{\hat{x}} D(\hat{x})\|_2 - 1)^2]$$

In this equation, the data distribution $P_{\hat{x}}$ is defined as a random interpolation of the generated data distribution and the real data distribution, while $\lambda$ is the penalty coefficient. In our experiments we found a value of 0.01 for $\lambda$ works well for our dataset. Utilising the RMSprop optimiser with a learning rate of 0.0001, the model's critic is trained five times for each training iteration of the generator. This training schedule assists in maintaining equilibrium between the two networks, promoting stable convergence. Due to the inherent randomness in model parameter initialization, and to ensure that the underlying distribution could be adequately captured, a total of 20 WGAN-GP models with distinctive parameter initialization were trained. Each model was subjected to 10,000 iterations to ensure convergence.

## 6. Non-dominant sorting optimisation-based generative adversarial network (NSGAN)

The proposed generative design framework, NSGAN, employs a non-dominant sorting multi-objective optimisation approach to search for novel samples with desired mechanical properties. Multi-objective optimisation problems (MOPs) represent a complex class of problems necessitating the simultaneous optimisation of multiple conflicting objectives. There are many methods have been developed to address MOPs, such as the weighted sum approach [48, 49] and goal programming [50], both of which tackle the multi-objective optimisation problem by transforming it into a single-objective optimisation problem. In addition, more advanced algorithms like the non-dominated sorting genetic algorithm (NSGA) and multi-objective particle swarm optimisation (MOPSO) [51] have been employed, where both these methods utilise the principle of Pareto dominance to identify non-dominated solutions. In the context of MOPs, one solution is said to dominate another if it is strictly better in at least one objective while being at least as good in all others [52]. Since achieving an optimal state simultaneously for all objectives in MOPs is typically impossible, the optimisation goal becomes to find a balanced set of non-dominated solutions which is referred to as the Pareto optimal set.

The NSGA-II, in particular, is prominent evolutionary algorithm characterised by its incorporation of a non-dominant sorting mechanism within the evolutionary optimisation procedure. This mechanism categorizes solutions in the population into various 'fronts' based on their dominance relationships and employs genetic operations such as selection, crossover, and mutation to iteratively evolve the population [34]. As an evolutionary algorithm, NSGA-II is capable of obtaining a set of solutions in a single run, without imposing assumptions on the mathematical properties of complex problems. This flexibility has led to its widespread application in solving MOPs across various fields [53-57]. It allows for a robust exploration towards the Pareto front while maintaining diversity among the solutions. While NSGA-II was selected for its proven efficacy and general applicability in a wide range of optimisation scenarios, it's important to note that this approach is not confined to this algorithm alone. It is adaptable to various other optimisation techniques, such as NSGA-III and MOEA/D, which may offer specific advantages in handling higher-dimensional objective spaces or more intricate Pareto front structures.

The architecture of the proposed NSGAN framework is illustrated in Figure 5.

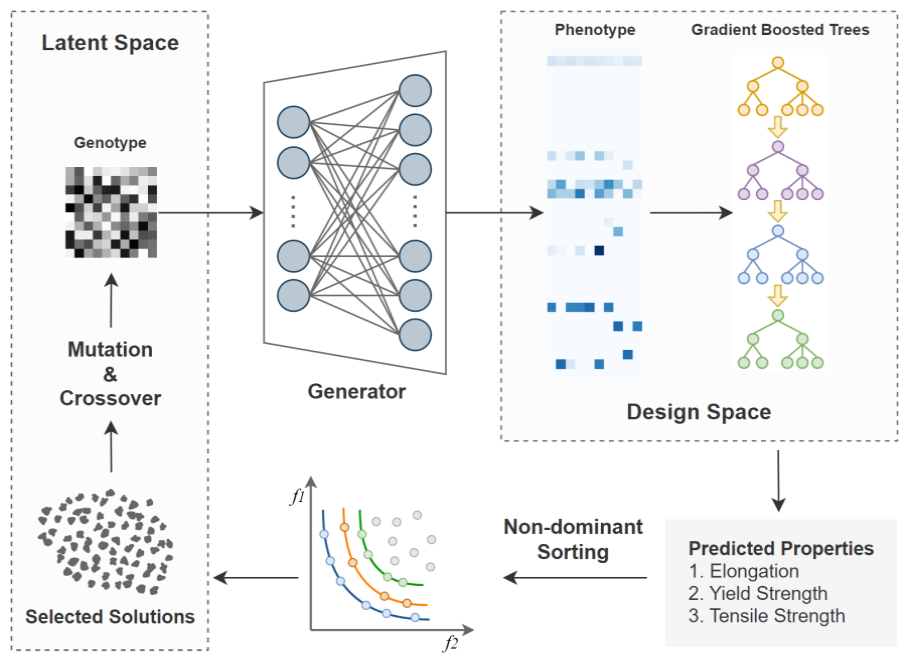

*Figure 5. Schematic representation of the proposed NSGAN framework. The genotype and phenotype are indicated by heatmaps, displaying the values of the latent variables and design specification across dimensions. The phenotype heatmap reveals that the data distribution in the design space is highly sparse, with most dimensions equal to 0.*

As previously noted, the proposed NSGAN framework applies the NSGA-II algorithm to search and optimise alloy samples, with the trained generator network functioning as a mapping that translates the latent space into the sparse design space. The optimisation starts with randomly generating a set of Gaussian distributed latent variables in the latent space, same as the GAN model's training process, which implies that the first generation (generation 0) of population corresponds to the original data distribution of the samples in the aluminium alloy dataset. The latent variables (genotypes) are decoded by the generator network into the design specification of the candidate solutions (phenotypes). Subsequently, using the pre-trained GBT models, the mechanical properties of the candidate solutions are predicted. Based on these predicted mechanical properties, the candidate solutions are processed through non-dominant sorting, grouping them into different Pareto fronts. The first front consists of all non-dominated solutions in the population. The second front includes all solutions dominated by the first front but not by others. This pattern continues for the subsequent fronts, each front represents a set of non-dominated solutions when compared to those in the succeeding fronts.

Table 3 provides the results across latent and design spaces of several solutions generated from an optimisation run using the NSGAN model. The genotypes in the latent space are represented by 10-dimensional vectors. These are then transformed by the GAN model's generator into 35-dimensional vectors (phenotypes), which include the molar ratios of 25 elements and 10 distinct processing conditions. For illustrative clarity in this presentation, here the phenotypes in the design space are depicted through their corresponding element compositions and processing conditions. Subsequently, the pre-trained GBT models are applied to predict the elongation and ultimate tensile strength (UTS) for these design candidates, which serve as the optimisation objectives for the NSGA-II model.

Based on the non-dominant sorting and crowding distance of the solutions, a subset from the population is selected for the evolutionary process. Crossover and mutation operations are then applied to the latent variables of these selected solutions to generate new candidates, leading to the formation of the next generation. The crossover operation combines the features of two parent solutions, while mutation introduces small random changes to a solution – with such an approach ensuring that the optimal solutions are carried forward to the next generation through the selection process, based on their Pareto efficiency and diversity. The process is repeated iteratively, gradually evolving the population towards the Pareto optimal set, while retaining a diversity among the solutions. The algorithm continues until specific termination criteria are met, such as achieving a preset goal or reaching the maximum number of iterations.

*Table 3.* NSGAN-generated non-dominated solutions mapped from latent space to design space with predicted mechanical properties.

| Latent space (10 dimensions) | Design space (35 dimensions) | | Predicted properties | |
|---|---|---|---|---|
| Latent variables | Element composition | Processing condition | Elongation (%) | UTS (MPa) |
| [-1.6, 0.9, 0.8, -2.2, 2.7, -1.4, 2.5, -2.8, 1.9, 2.1] | $Al_{0.852}Cu_{0.0135}Mg_{0.0245}$ $Sc_{0.0022}Zn_{0.107}Zr_{0.0009}$ | Solutionised + artificially peak aged | 9.40 | 775.43 |
| [-1.9, 0.9, 1.7, -2.2, 2.7, -1.2, -1.7, -2.8, 2.2, 2.1] | $Al_{0.849}Cu_{0.061}Fe_{0.0057}$ $Mg_{0.011}Ti_{0.0012}Zn_{0.07}$ | Solutionised + cold worked + naturally aged | 18.40 | 623.24 |
| [0.8, 0.6, 2.7, -0.9, -1.6, 1.8, -2, 2.5, -0.2, 1.2] | $Al_{0.955}Cu_{0.038}Li_{0.0057}$ $Zr_{0.0017}$ | Solutionised + naturally aged | 23.10 | 499.04 |
| [-2.5, -1.6, -2.4, -1.2, -0.04, -2.1, -0.7, -2.7, -2.2, -1.9] | $Al_{0.943}Mg_{0.056}Mn_{0.0013}$ | No processing | 28.87 | 321.82 |
| [-1.7, -0.5, 1.7, 2.1, 2.9, -1.6, 2.2, 2.6, -0.3, 0.5] | $Al_{0.9994}Cu_{0.0005}Fe_{0.0001}$ | No processing | 44.24 | 69.94 |

The NSGA-II algorithm was implemented using Pymoo [58] which is a python library known for its comprehensive collection of state-of-the-art multi-objective optimisation algorithms and quality indicators. Its modular structure allows for easy customisation and extension, making it a suitable choice for the present application.

In configuring our model, binary tournament selection was employed as the selection method. The crossover probability was set at 0.9, with a mutation probability of 0.1. The population size was fixed at 200 to maintain diversity. Our observations indicated that the model usually starts converging within 100 iterations; therefore, the entire process was designed to run for a total of 500 iterations to facilitate appropriate convergence.

## 7. Web application and user tool

In this study, an online tool applies the proposed NSGAN framework was developed based on Streamlit, which is an open-source Python library expressly designed to create interactive web applications. This online tool enables users to manually input element composition and processing condition to predict mechanical properties of aluminium alloys. And by simply setting parameters such as population size and the number of generations, users can apply the proposed multi-objective evolutionary optimisation procedure to generate a range of solutions optimised for strength and ductility and predict their mechanical properties. While the online tool is devised based on models trained with the aluminium alloy dataset, by integrating the NSGAN framework with various databases, it could set the stage for broader applications in diverse domains. The web applications for property prediction and NSGAN model can be accessed at [Aluminium Alloy Property Prediction (https://aluminium-alloy-property-prediction-app.streamlit.app/)](https://aluminium-alloy-property-prediction-app.streamlit.app/) and [NSGAN Generative Aluminium Alloy Design (https://nsgan-generative-aluminium-alloy-design.streamlit.app/)](https://nsgan-generative-aluminium-alloy-design.streamlit.app/), respectively. This serves as a demonstration of the potential of the NSGAN generative framework to be a foundational tool, for the design and exploration of novel materials in general. It also provides an interactive user tool that is relevant to alloy designers and alloy practitioners.

## 8. General discussion

The proposed NSGAN framework has been utilised and explored in several experiments to discover the correlation between the latent space and design space; and to trace the continuous evolution of the data distribution in design space during the multi-objective evolutionary optimisation process. The efficacy and efficiency of the proposed model was showcased via a comparative analysis of the proposed NSGAN framework in relation to two prevalent methods in alloy design. Through this generative procedure, a substantial number of novel alloys were generated and compared with samples from the training set, from which several clusters representing potential superior performance alloys are recognised.

### 8.1. Gene correlation

The generator network of the GAN model serves as a decoder that processes latent variables (the genes) through a complex series of nonlinear mathematical operations into synthetic alloy samples within the design space. Given that the GAN model's training involves random sampling in latent space, the resultant mapping by each trained generator network exhibits high nonlinearity and randomness.

To visualise this complex mapping, a trained generator network was utilised to traverse each dimension in the latent space within a range of [-3, 3], encompassing the range of 99.73% of randomly sampled values. The mapping is illustrated through the variation of design specifications corresponding to changes in the latent variables. Due to the high dimensionality of the design space, we selected two elements that are respectively abundant and scarce within the database for comparison. Heatmaps were used to illustrate their variations corresponding to changes in the value of each gene, depicted through shifts in colour depth. To better visualise the variation in the element's molar ratio concerning individual genes, we plotted the average element molar ratio for each gene in conjunction with changes across the other nine dimensions. Figure 6 illustrates the correlation between ten dimensions in the latent space and the molar ratios of aluminium (Al) and silver (Ag), where the correlation between each dimension and the molar ratio of elements in the generated alloy samples reveals a highly nonlinear transformation. And it can be observed that the generated alloy samples tend to possess a higher molar ratio of aluminium with less minor elements when the input neurons of the generator network are deactivated. Furthermore, as shown in Figure 6(b), minor elements such as Ag exhibit a very sparse distribution within the design space, with the molar ratio being zero across the majority region. This occurrence is attributed to the fact that the minor elements appear only in a subset of the training data, which is also the cause of the overall sparsity in the design space.

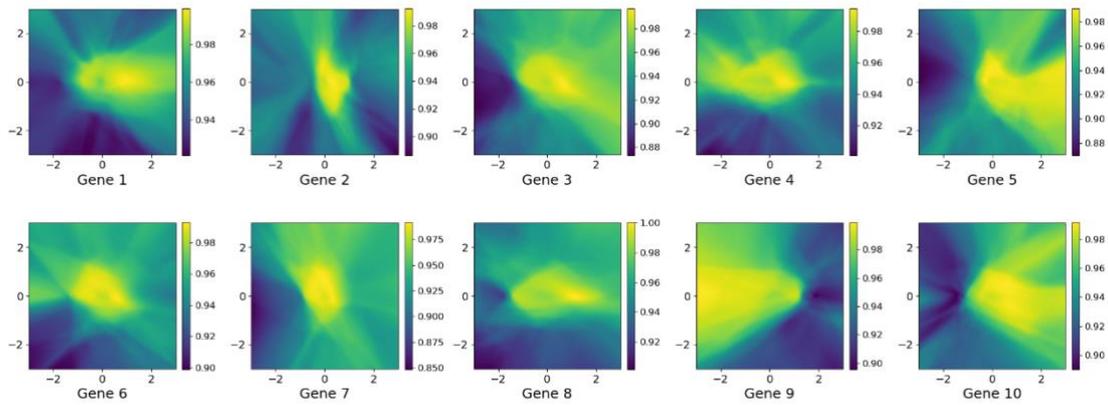

(a)

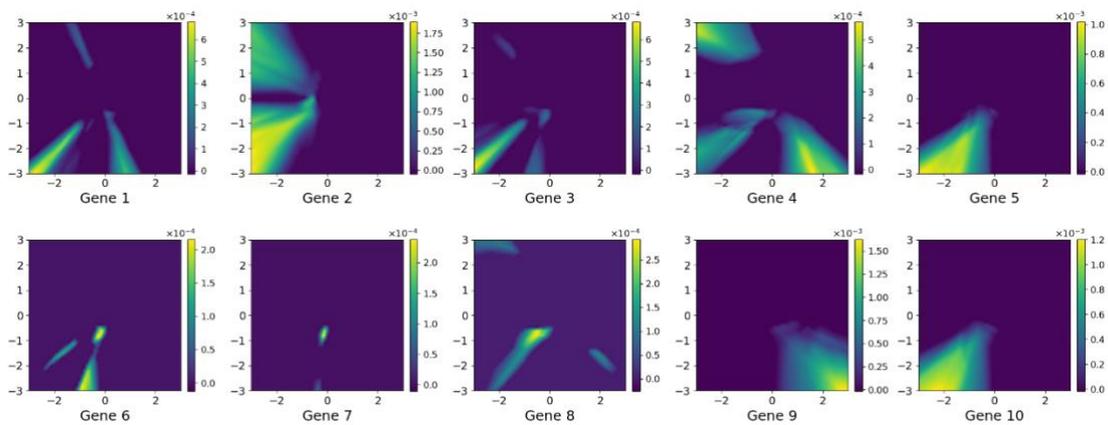

(b)

***Figure 6.*** *Visualisation of the molar ratio variations of (a) aluminium, and (b) silver, across latent space dimensions.*

Additionally, by plotting the variations between the processing condition and latent variables, we observed that each latent variable typically correlates with two primary processing conditions. Figure 7 illustrates the variations in the synthetic alloy sample's processing condition with respect to the value of gene 1 and gene 3. It can be seen that a single gene may exhibit a relationship that is nearly binary with the processing condition. For example, for gene 1, the generated alloy samples' processing condition is consistently 'No processing' when greater than 0.5, whereas it is 'Solutionised + Artificially peak aged' for values less than 0. Gene 3, on the other hand, primarily correlates with 'Solutionised + Artificially peak aged' and 'Strain Hardened (Hard)'. This may be due to the fact that the processing conditions, as one-hot-encoded categorical data, have a relatively simple distribution. It is noted that this pattern is only evident when observing a single gene variation, and the relationship becomes highly nonlinear when multiple genes change simultaneously.

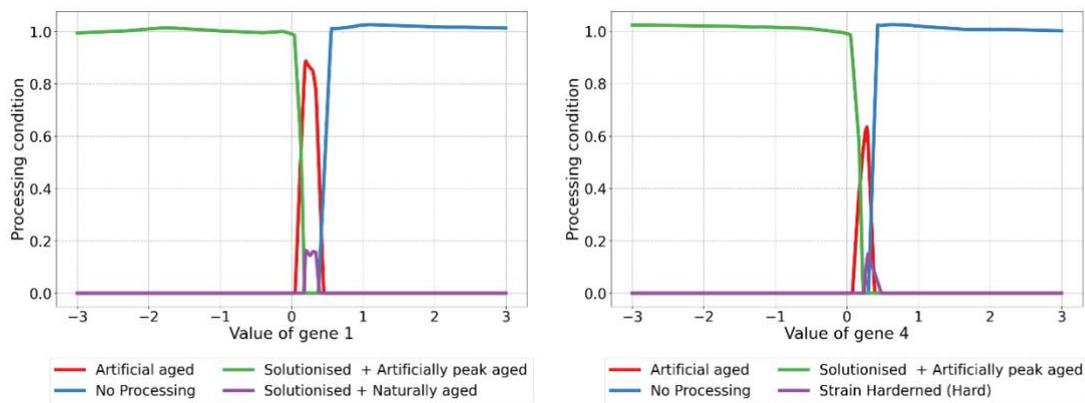

*Figure 7.* Variation in processing condition concerning the values of gene 1 and gene 4. Only processing conditions that have non-negative values are shown; the y-axis represents their probability distribution.

The above analysis reveals that the correlation between the latent space and the design space is characterised by a (highly) nonlinear relationship. While some patterns are discernible, such as the activation of latent variables increasing the content of minor elements and the binary correlations between individual genes and processing conditions, the complexity of this relationship underscores the challenges in using the GAN model alone to perform targeted generation for alloys with specific elements or properties.

*8.2. Data distribution evolution*

To facilitate a more intuitive understanding of the multi-objective design process, we conducted a study to monitor and visualise the data distribution of the population throughout the evolutionary optimisation procedure. Recognising that the changes in data during the latter stages of optimisation are subtle and challenging to depict, we focused our attention on the visual representation of the earlier generations within the procedure. Figure 8 presents the t-SNE dimension-reduced representations of the population at Generations 0, 2, 5, 20, 100, and 200, with the density distribution of the generator network elucidated by the density heatmap in the background. This visualisation reveals the evolution and variety of alloys within the population across generations. The latent variable of the population in Generation 0 are randomly sampled by a multi-variate Gaussian distribution, same as in the GAN model's training, which results in the population at Generation 0 essentially aligns with the density distribution of the generator network, corresponding to the original training set's distribution. As the optimisation progresses, the non-dominated samples with exceptional strength-ductility trade-offs are selected to form successive generations, leading to a gradual convergence of the population to specific regions, as depicted in Figure 8. Notably, from around Generation 20 and onwards, the regions of convergence are observed within the less dense regions of the generator's density distribution, with some of the samples distributed along the margins, which are regions that the generator network unlikely to generate during random sampling. Furthermore, a comparison between Generations 100 and 200 reveals a more subtle distribution shift within the population, illustrating the algorithm's focus on increasingly refined optimisation within certain alloy systems as the iterative optimisation process progresses. These comparative visuals also demonstrate that, despite the targeted optimisation, the population maintains a degree of diversity. Conclusively, the optimisation process evolves from random sampling based on the GAN model's distribution to a more targeted and iterative optimisation in particular regions of this distribution, which underscores the framework's effectiveness in navigating and fine-tuning the design space within the given alloy systems.

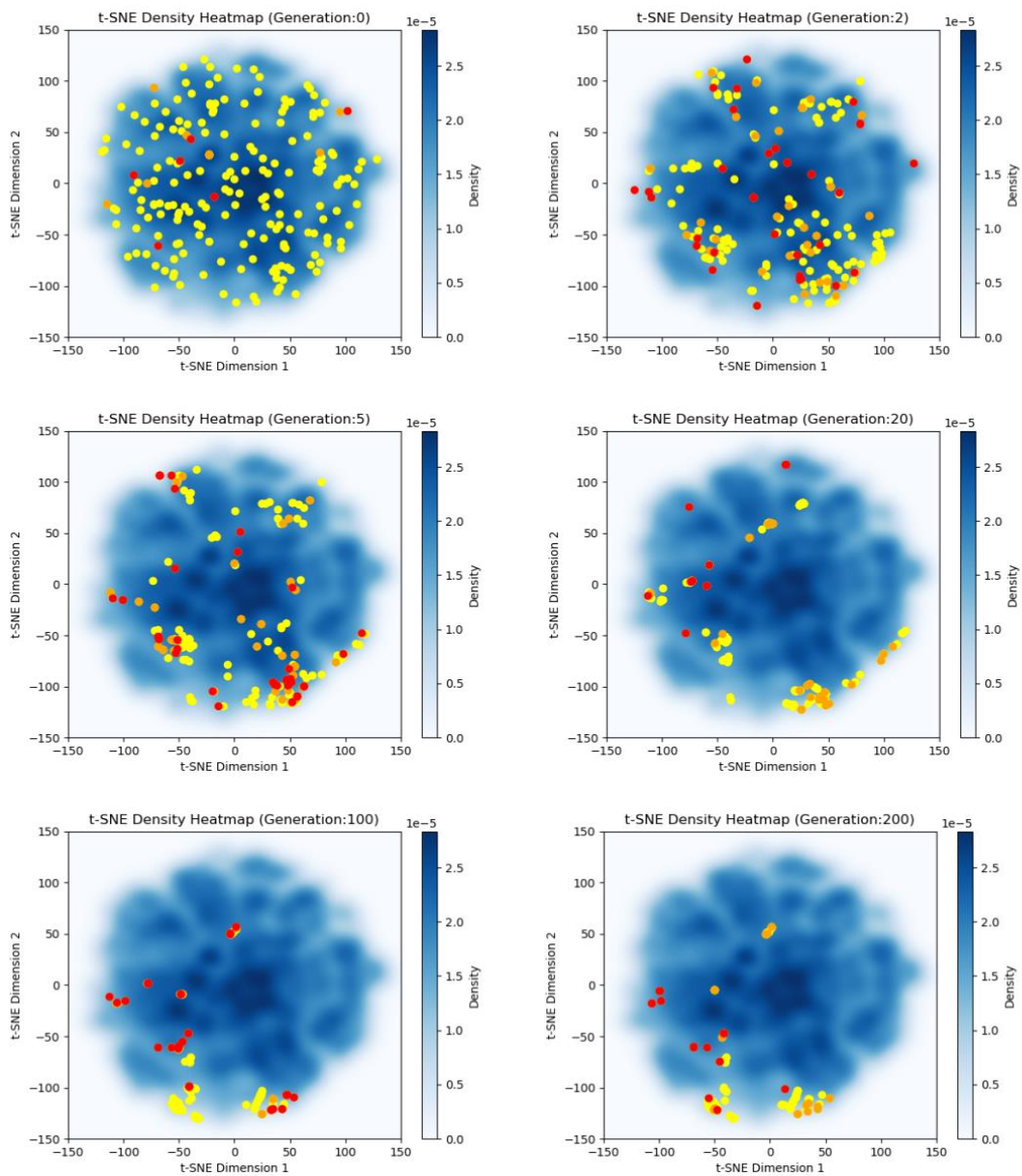

*Figure 8.* *The comparison between the GAN generated data distribution and the population for different Generations. The population is depicted with yellow points, with the first Pareto front highlighted in red and the second Pareto front in orange.*

Figure 9 presents the scatter plot of the predicted tensile strength and elongation for the non-dominated solutions across difference Generations. Given that data changes are subtle during the later stages of optimisation, only Generations 0, 2, 5, and 20 are presented, which ensures the illustration of the figure remains clear and effective. The plot reveals a gradual discovery of solutions with superior strength-ductility trade-offs in the first 20 generations, particularly in the range where Tensile Strength exceeds 250 MPa. In this context, an elite strategy is employed to ensure the retention of optimal solutions by preserving the non-dominated solutions from each preceding generation. This method effectively aids in better illustrating and visualising the progressive improvements throughout the optimisation process. As a result, throughout the optimisation process, the model is observed to continually explores and advances towards the Pareto front as the number of Generations increases.

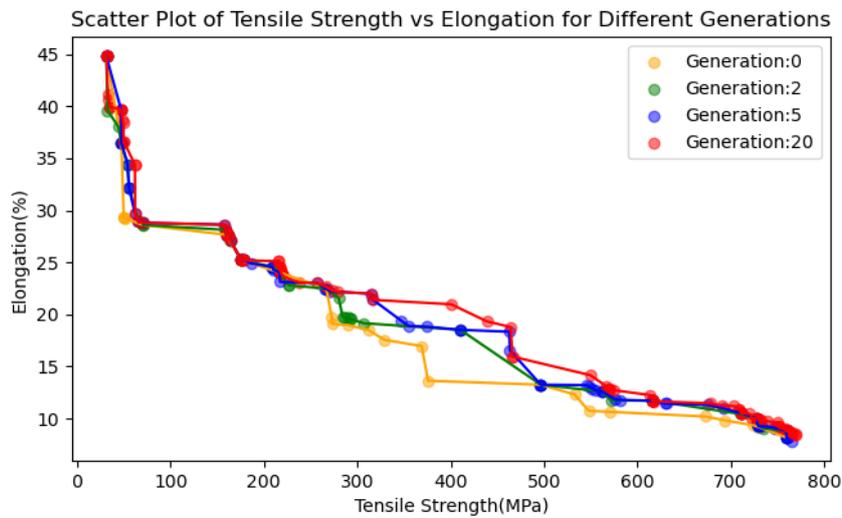

*Figure 9.* The scatter plot of Tensile Strength vs Elongation of the non-dominated solutions for different Generations (right).

The average novelty of the population across different generations are depicted in Figure 10. The novelty of a generated alloy sample is measured by the minimum Euclidean distance to the existing alloy samples within the training set. As observed from Figure 10, the average novelty experiences a significant increase in the initial ten generations, followed by a gradual decline, before it once again begins to rise steadily after approximately 70 generations. This trend can be attributed to that, in the first 10 generations, the genetic operations are conducted on alloy candidates broadly distributed in the design space, which allows alloy samples from various regions to be 'crossed over' (as indicated by the early diverse distribution in Figure 8), leading to the generation of a considerable novelty. Many of these novel samples, however, are dominated solutions. Therefore, as the optimisation continues, a significant portion of these novel samples are gradually phased out, resulting in a decline in the average novelty. In the later stages of the optimisation process, the data distribution in the population continues converging within the design space. Through iterative genetic operations, the model gradually explores increasingly novel solutions that possess optimal properties. Consequently, the average novelty of the population begins to rise steadily, reflecting the continued exploration and refinement within the design space. This observation aligns with our aim for the evolutionary optimisation to facilitate the exploration of novel optimal alloy samples.

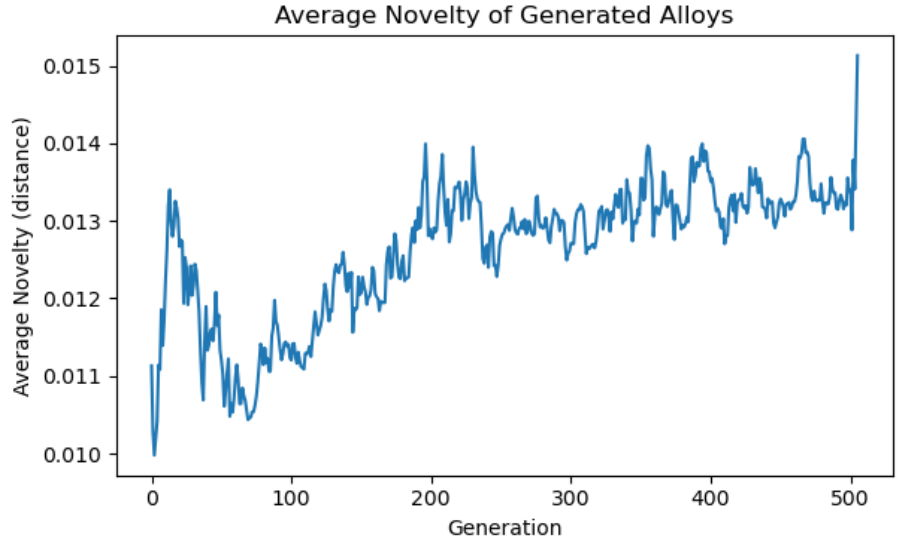

*Figure 10.* *The average novelty of the population in the 500 generations.*

*8.3. Model comparison*

In this section, we conduct a comparative analysis of the proposed NSGAN framework in relation to two prevalent methods in alloy design. The first method encompasses data-driven generative machine learning models [13-15], with our focus on GANs as a representative example, hereafter referred to as the GAN method. The second approach adopts optimisation-centric, ML-supported strategies, such as the integration of GAs with property prediction ML models for alloy design, as explored in the literature [22-24]. For this comparison, we employ the NSGA-II algorithm, designated as the GA+ML method.

*8.3.1 NSGAN vs. GAN*

GANs are well acknowledged for efficiency in generating synthetic samples via learning from existing datasets. When integrated with property prediction ML models, constraints can be imposed on GANs to target specific property optimisations in a desired direction [23, 24]. However, GAN-generated samples often exhibit considerable randomness and generally lack controllability. This randomness, combined with the tendency of the majority of samples to closely resemble those in the original dataset, necessitates the generation of a large volume of samples to adequately search the design space. Subsequently, the process involves sifting through the large volume of randomly generated samples to identify and evaluate those demonstrating potentially superior performance and novelty. Unfortunately, such samples typically constitute a very small proportion of the total generated, posing a significant challenge in utilising GANs for efficient and targeted material design.

Herein, 20 WGAN-GP models with distinct parameter initialization were trained and applied to compare the performance of direct GAN-generated alloy samples against samples generated using the proposed NSGAN framework. In the GAN approach, each WGAN-GP model was employed to generate 50,000 alloy samples, culminating in a total of 1,000,000 samples across all models. The NSGAN method involved running the NSGAN framework once with each WGAN-GP model, with a population size of 200 over 500 generations, leading to a collective output of 4,000 alloy samples from all models.

*Table 4. Comparative results of GAN and NSGAN methods in generation performance.*

| Model | Data generated | Average elongation | Average UTS | Samples within pre-defined criterion | Average novelty |
|---|---|---|---|---|---|
| GAN | 1000000 | 13.15 % | 317.4 MPa | 116 (0.01%) | 0.0096 |
| GA+ML | 4000 | 31.32% | 302.73 MPa | 563 (14.07%) | 0.0199 |
| NSGAN | 4000 | 18.97 % | 453.91 MPa | 163 (4.08%) | 0.0145 |

The comparative results of these two methods are presented in Table 4, which includes the average UTS, average elongation, average novelty (measured by the Euclidean distance to training data), and the proportion of samples that meet a pre-defined criterion: UTS > 600 MPa and elongation > 15%. From the table, it is evident that the NSGAN model has generated samples with notably higher average elongation (18.97%) and average UTS (453.91 MPa) compared to the GAN model. Moreover, despite the significantly lower total number of samples generated by the NSGAN model compared to the GAN model, the NSGAN model still produces a similar number of samples that meet the pre-defined criteria. In terms of percentage, within the results generated by NSGAN, 4.08% meet the predefined criterion, which is significantly higher compared to the mere 0.01% achieved through direct GAN generation. This indicates the NSGAN model's higher efficiency in generating samples with the desired mechanical properties, showcasing its superior data utilisation and targeted search capabilities in the design space. Furthermore, the average novelty score of the NSGAN model is higher (0.0145) than that of the GAN model (0.0096). This implies that the NSGAN model's outputs are potentially more diverse and innovative compared to the GAN-generated samples, which is a desirable feature in exploring novel alloy compositions. In summary, the NSGAN model demonstrates a more targeted and efficient approach to generating novel high-performance alloy samples.

*8.3.2 NSGAN vs. GA+ML*

The GA+ML method operates by setting specific ranges for parameters within a designated search space. It leverages ML to predict the performance of samples, thereby identifying potential candidates with superior characteristics. As previously mentioned, unlike data-based methods, this approach enables effective exploration of novel designs, particularly excelling in scenarios characterized by low dimensionality and robust accuracy of ML model predictions. However, when confronted with complex, high-dimensional challenges, this method often encounters issues related to significant discrepancies between generated results and training data. Such discrepancies can undermine the reliability of ML predictions, rendering the outcomes of the iterative optimisation process ineffective.

For the GA+ML method parameter optimisation was based on the molar ratio upper and lower limits of 657 alloy samples from the training set, as illustrated in Table 1. To maintain consistency in our comparative analysis, we also employed an identical setup of a population size of 200 and 500 iterations across 20 runs, resulting in a total of 4000 generated samples.

From the results in Table 3, it is evident that the GAML method yields a commendable performance, with 563 samples meeting the criterion of UTS > 600 MPa and elongation > 15%, accounting for 14.07% of the total number of generated samples. However, a closer examination of the generated results indicates the presence of potential erroneous predictions.

As illustrated in Table 5, regarding processing methods, the GA+ML method's results are constrained to a mere three processing methods. Furthermore, it predominantly recommends 'Naturally aged' as the processing method for all alloys with medium and high strength (UTS > 600 MPa), a recommendation that is at odds with established metallurgical practices [59]. In contrast, the NSGAN results display a diverse range of processing methods that correspond with traditional metallurgical expertise. This lack of diversity and the deviation from established practices seen in the GA+ML results highlight the issues associated with high-dimensional search spaces. The limited predictive ability of an ML model, trained on a constrained set of existing alloy samples, can result in erroneous predictions for randomly generated samples within the high-dimensional space, ultimately guiding the algorithm to converge on invalid solutions.

*Table 5. Comparative distribution of processing methods for alloy samples by GA+ML and NSGAN approaches.*

| Processing method | GA+ML | | NSGAN | |
|---|---|---|---|---|
| | Overall | UTS > 600 | Overall | UTS > 600 |
| No processing | 2487 | 0 | 1701 | 0 |
| Solutionised + Artificially peak aged | 0 | 0 | 1111 | 806 |
| Solutionised + Naturally aged | 939 | 0 | 763 | 91 |
| Solutionised + Artificially over aged | 0 | 0 | 254 | 164 |
| Naturally aged | 574 | 563 | 96 | 17 |
| Solutionised + Cold Worked + Naturally aged | 0 | 0 | 26 | 26 |
| Artificial aged | 0 | 0 | 19 | 13 |
| Strain hardened | 0 | 0 | 18 | 14 |
| Strain Hardened (Hard) | 0 | 0 | 12 | 0 |
| Solutionised | 0 | 0 | 0 | 0 |

When examining the elemental composition aspect of the outcomes generated by the GA+ML method, as illustrated in Table 6, it is observed that for alloys with medium to high strength—the most 'valuable' ones—approximately 97.5% of the samples generated by the GA+ML method feature heavy additions of Zn, exceeding 10% Zn content. Furthermore, a significant 33.7% of these samples contain more than 15% Zn. Such high concentrations are typically challenging to realize in practice due to the propensity for "hot cracking" during solidification [60] with practical limits of ~12 wt.% only achieved by novel spray forming [61].This tendency may stem from the ML model's recognition of a strong positive correlation between Zn content and strength within the training data, leading to a biased generation of high-Zn samples erroneously predicted to have superior performance.

*Table 6. Zinc concentration fraction in high-strength alloys generated by GA+ML and NSGAN models.*

| Model | Count UTS > 600 | Average Zn fraction (wt. %) | Count Zn > 10% | Count Zn > 15% |
|---|---|---|---|---|
| GA+ML | 563 | 0.142 | 549 | 190 |
| NSGAN | 1131 | 0.099 | 470 | 32 |

On the other hand, the NSGAN method generates samples based on existing alloy compositions, whose distribution inherently incorporates the empirical knowledge of conventional alloy design. This approach, therefore, manages to circumvent the limitations posed by the restricted generalization capability of ML models trained on limited data, avoiding the skewed emphasis on Zn observed in the GA+ML method's results.

The emergence of such anomalies within the GA+ML method can be traced to the intrinsic limitations of ML models; these models exhibit commendable predictive prowess within the bounds of similarity to their training datasets but falter when extrapolating significantly beyond these confines. Considering the sparse nature of the training data's manifold within the expansive high-dimensional design space, the probability of intersecting this manifold during the stochastic search process employed by the GA+ML method is notably small. This accounts for the pronounced deviation of the GA+ML results from the established patterns of the training data. Most of these results bear apparent discrepancies and do not seem to have assimilated the fundamental principles manifest in the training alloy designs.

The NSGAN model, initiating its search from the distribution of existing samples, navigates the design space in proximity to this distribution, allowing for the generation of novel yet familiar designs. The search initiation, being cantered around the training data's distribution, enables the ML models to yield predictions with enhanced reliability, particularly in the initial stages of optimisation. Furthermore, as evidenced in Figure 8, even after convergence, the distribution of data within the population remains closely situated around the training data's distribution, without significant divergence. Thus, the resultant designs predominantly encapsulate the 'essence' of human-led alloy design, reflecting the expertise and knowledge learned from the training dataset. This correlation between the generated compositions and the training data distribution underpins the NSGAN's ability to innovate while retaining the foundational aspects of alloy design.

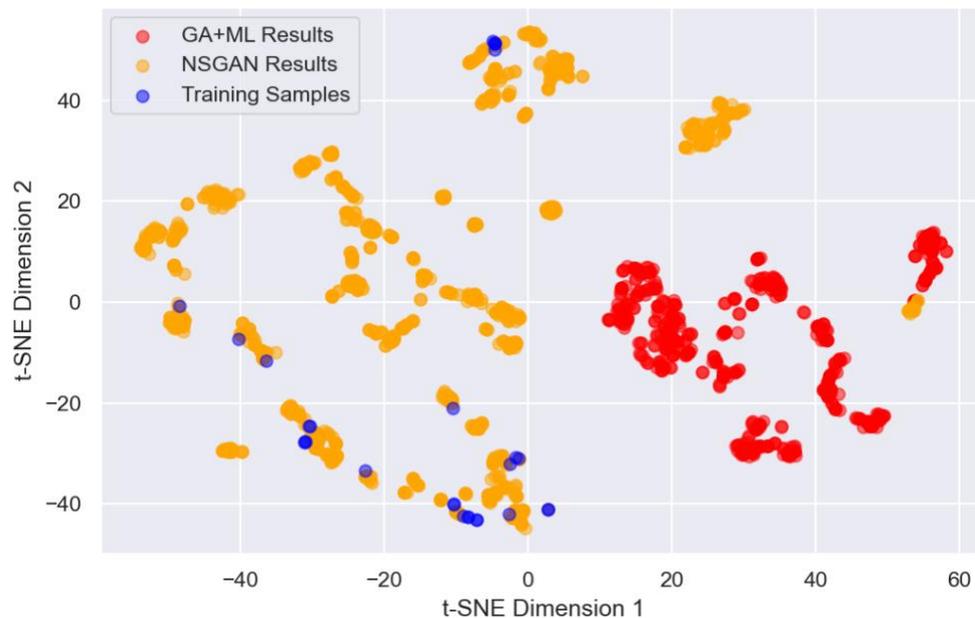

*Figure 11.* t-SNE representation of high-strength alloy samples from GA+ML and NSGAN methods compared to training samples.

To visualise the data distribution of the generated results, t-SNE was utilised to project the high-strength alloy samples (UTS > 600 MPa) onto a two-dimensional plane as seen in Figure 11. The resulting scatter plot reveals that the GA+ML results (red points) are predominantly clustered in a separate area from the training samples (blue points), suggesting a significant divergence in the feature space. In contrast, the NSGAN results (yellow points) exhibit clusters that overlap with the training samples to some extent, indicating a degree of similarity. This overlap implies that the NSGAN framework could balance exploration and exploitation, generating novel alloy designs that adhere to the empirical data distribution.

*8.3.3 Computation time and efficiency*

When comparing the NSGAN with the GA+ML method under the identical setup of a population size of 200 and 500 iterations across 20 runs, the NSGAN model clocked in at approximately 14.5 seconds for each run, while the GA+ML method required about 24 seconds. This observation underscores the NSGAN model's notable efficiency, even though it integrates an additional decoding step from the latent space to the design space. This efficiency gain is likely attributable to the dimensional reduction feature inherent to the NSGAN framework. In contrast, the GA+ML method operates within the original high-dimensional space, where the enlarged search space increases the computational effort required at each step of the optimisation process.

The GAN method, pre-trained for direct sample generation, bypasses the iterative optimisation process, theoretically offering rapid sample production post-training. However, application of the GAN for high-performing novel sample generation necessitates producing a vast array of samples, followed by the meticulous process of selection and identification of optimal solutions, which results in computation time varying significantly depending on the design requirements. For instance, in the prior experiment where one million samples were generated, only 116 satisfied the pre-defined selection criteria. Generating those one million alloy samples required ~6 minutes, while the subsequent statistical analysis to filter out the optimal samples took an additional ~7 minutes. In this light, although the precise computational time for employing the GAN method is indeterminate, it is evidently higher compared to both the NSGAN and GA+ML methods. The experiments herein were conducted on a computer equipped with the following specification: CPU i9-13900K 3.00 GHz and 32GB of RAM.

In the comparative analysis of the NSGAN framework against the GAN and GA+ML methodologies, it is evident that the proposed NSGAN method presents several advantages. Notably, NSGAN distinguishes itself through its remarkable efficiency, demonstrated by reduced computation times and its ability to generate fewer samples yet achieve a higher proportion of samples possessing desired mechanical properties. Its dimensionality reduction feature significantly conserves computational resources throughout the iterative optimisation process. Furthermore, NSGAN demonstrates an exceptional ability to balance innovation with the prediction reliability. By employing GAN models to internalise the empirical knowledge found within existing alloy samples of conventional alloy design, NSGAN enables a more focused and effective search and optimisation for novel samples. In summary, it presents a method for a more comprehensive and pragmatic approach to computational alloy design, positioning itself as a superior alternative in terms of both efficiency and efficacy.

*8.4 Generated results*

Utilising the trained GAN models and employing the proposed NSGAN framework, a total of 3554 non-dominated aluminium alloy samples optimised for strength and ductility were explored. The results are illustrated in Figure 12, which presents a scatter plot of Tensile Strength vs Elongation for all the generated optimal samples and the existing samples from the training set. It can be observed that the mechanical properties of most of the generated optimal samples align closely with the 'Pareto front' of the training data. Within this data scattering, several clusters can be identified that may represent potential exemplars of superior performance, as highlighted by the regions encircled in green in the figure. Some examples of these alloy samples are detailed in Table 7. The complete dataset can be found in the supplementary file, accessible via the following (link for Google Doc).

*Table 7. Example of some novel aluminium alloys explored by the proposed NSGAN framework, including their alloy composition, processing method, and predicted mechanical properties.*

| Alloy composition | Processing condition | Tensile strength (MPa) | Yield strength (MPa) | Elongation (%) |
|---|---|---|---|---|
| $Al_{0.8568}Cu_{0.0125}Mg_{0.0256}Mn_{0.0007}Zn_{0.1013}Zr_{0.003}$ | 8 | 779.64 | 716.95 | 16.97 |
| $Al_{0.8182}Cr_{0.005}Cu_{0.0148}Mg_{0.0271}Zn_{0.1328}Zr_{0.0022}$ | 8 | 778.17 | 738.45 | 17.38 |
| $Al_{0.8296}Cr_{0.0048}Cu_{0.0123}Mg_{0.0236}Zn_{0.1273}Zr_{0.0024}$ | 8 | 777.89 | 736.11 | 17.49 |
| $Al_{0.8519}Cu_{0.0157}Mg_{0.0212}Ti_{0.0003}Zn_{0.107}Zr_{0.0038}$ | 8 | 765.59 | 704.03 | 17.10 |
| $Al_{0.863}Cr_{0.005}Cu_{0.026}Fe_{0.005}Mg_{0.024}Mn_{0.0014}Ti_{0.001}Zn_{0.076}$ | 2 | 635.44 | 574.74 | 20.87 |
| $Al_{0.851}Cu_{0.06}Fe_{0.007}Mg_{0.0016}Mn_{0.002}Ni_{0.005}Ti_{0.0006}Zn_{0.072}$ | 2 | 623.63 | 563.28 | 22.80 |

*\* Processing condition 2: Naturally aged*      *\* Processing condition 8: Solutionised + Naturally aged*

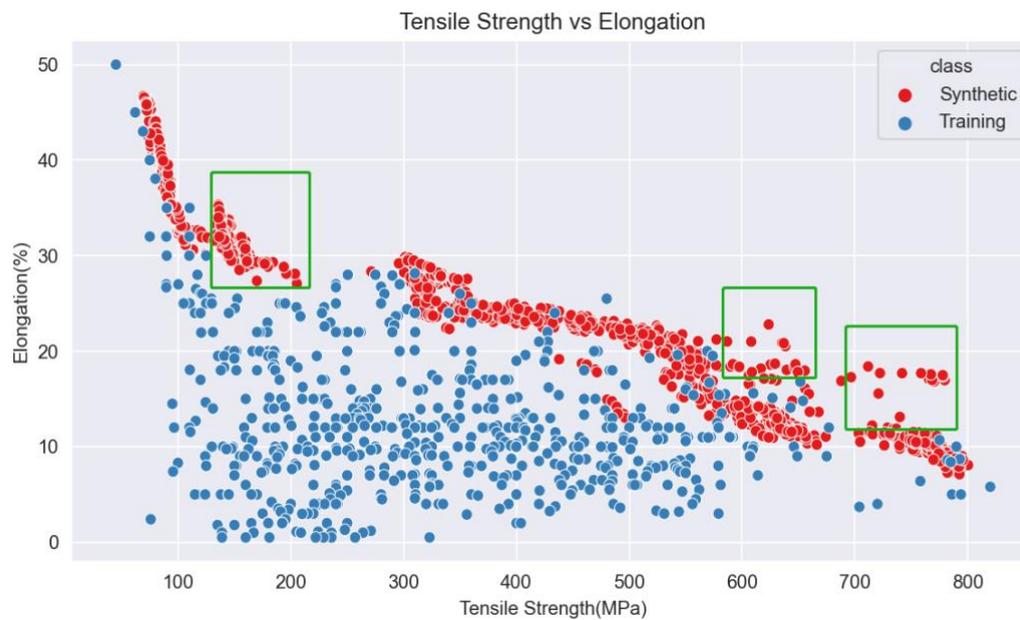

*Figure 12. The scatter plot of Tensile Strength vs Elongation of all the non-dominated solutions generated by 20 different generator models.*

## 9. Conclusions and future works

In this research, a novel generative design framework was introduced, called NSGAN (non-dominant sorting optimisation-based generative adversarial network); formulated for the exploration and high throughput innovation design of materials with desirable properties. The synergy of GAs and GANs has been harnessed, exploiting their respective strengths, and mitigating their weaknesses, to facilitate the design and optimisation of new alloys.

The NSGAN framework employs GANs to overcome the complex "curse of dimensionality" problem, simplifying the high-dimensional sparse data distribution into a more manageable lower-dimensional space. This innovation enhances the efficiency of the GA search process. The combination of GA and GANs directs the generative process towards creating innovative samples with targeted attributes. The alignment of the generated sample distribution with the original training data ensures the reliability of the predictive property ML models. Utilising a comprehensive dataset of 657 aluminium alloys, the efficiency and efficacy of the approach was demonstrated. The framework addresses prevalent conflicting objective optimisation problems, like the strength-ductility trade-off in aluminium alloys, by employing a multi-objective optimisation evolutionary algorithm, NSGA-II. To promote interaction with the research, an online tool that integrates the proposed model was created; offering a platform for other researchers or practitioners to apply the design framework for customised material design requirements.

While the NSGAN framework introduced in this research demonstrates an exceptional ability in balancing exploration and exploitation, its computational nature results in certain limitations at this stage. The validity and accuracy of this approach ultimately require empirical experimental validation. Future work will focus on extensive experimental validation to confirm the accuracy of mechanical property predictions for new alloy samples generated by the NSGAN. This includes, but is not limited to, testing key performance indicators such as ultimate tensile strength and ductility against those of existing alloy samples. Incorporating high-throughput experiments within the generative design cycle serves to elevate the performance and precision of the proposed generative framework. Additionally, future research will also explore the extension of the NSGAN framework to other material systems, verifying its adaptability and flexibility across diverse material design challenges.

Overall, the proposed NSGAN framework represents an innovative intersection between material science and computational techniques, aligned with the emerging paradigm of the material genome. It provides both a theoretical foundation and a practical methodology, which could be considered for extension to other domains of material exploration. The interdisciplinary nature of this work illustrates a possible direction for future studies and offers an encouraging example of how innovative approaches might influence the field of material design.

**Data availability**

The raw data used in this study is openly available at the following repository: https://doi.org/10.17632/b6br4yk6r3.1.

**Code availability**

The codes developed in this study are openly available on GitHub at the following repository: https://github.com/anucecszl/NSGAN_aluminium.

**Acknowledgements**

We gratefully thank Ninad Bhat for discussions.

**Author contributions**

ZL performed the coding and development of the models. Both NB and ZL contributed in the study design and preparation of the manuscript. NB supervised the study and lead PI.

**Competing interests**

The authors declare no competing interests.